\documentclass[12pt,a4paper]{article}
\usepackage{cite}
\usepackage{graphicx}
\usepackage{amssymb}
\usepackage{epsfig}
\usepackage[tmargin=1cm,bmargin=1cm,lmargin=2cm,rmargin=2cm]{geometry}

\nopagebreak

\begin{document}
\pagestyle{empty}
\baselineskip 22pt

\begin{flushright}
\end{flushright}
\vskip 65pt
\begin{center}
{\Large \bf Unparticles in diphoton production to NLO in QCD at the LHC}

\vspace{8mm}
{ \bf
M. C. Kumar$^{a,b}$
\footnote{mc.kumar@saha.ac.in},
Prakash Mathews$^a$
\footnote{prakash.mathews@saha.ac.in},
V. Ravindran$^c$
\footnote{ravindra@mri.ernet.in},
Anurag Tripathi$^c$
\footnote{anurag@mri.ernet.in}
}\\
\end{center}
\vspace{10pt}
\begin{flushleft}
{\it
a)
Saha Institute of Nuclear Physics, 1/AF Bidhan Nagar,
Kolkata 700 064, India.\\

b)
School of Physics, University of Hyderabad, Hyderabad 500 046, India.\\

c)Regional Centre for Accelerator-based Particle Physics, 
Harish-Chandra Research Institute,
 Chhatnag Road, Jhunsi, Allahabad, India.\\
}
\end{flushleft}
\vspace{10pt}
\begin{center}
{\bf ABSTRACT}
\end{center}

We compute to next-to-leading order in QCD the tensor unparticle 
contribution to the diphoton production at the LHC, wherein the 
unparticle sector is a consequence of (a) scale invariance 
but not full conformal invariance and (b) conformal invariance.  
We use the semi-analytical two cutoff phase space slicing method 
to handle the ${\cal O}(\alpha_s)$ corrections to the $p ~p \to 
\gamma ~\gamma ~X$ and show that our results are insensitive to the 
soft and collinear cutoffs.  In order to avoid the contribution 
of the photons due to fragmentation, we employ the smooth cone 
isolation criterion.  Significance of the QCD corrections to the 
diphoton events including unparticles is highlighted.

\vspace{1cm}
\rightline{PACS numbers: 12.60.-i, 11.25.Hf, 14.80.-j}
\vskip12pt
\vfill
\clearpage

\setcounter{page}{1}
\pagestyle{plain}

\section{Introduction}
Diphoton signals play very important role in the phenomenological studies 
of the standard model (SM) and also of beyond the standard model (BSM).  
These photons can constitute a potential background to those coming from 
Higgs decay.  In the BSM, namely, supersymmetric \cite{Nojiri:2008}, 
extra-dimensional 
\cite{Eboli:2000} and unparticle (UP) models \cite{Kumar:2008zz} the diphoton 
productions can constrain the model parameters.  Hence it is important 
to have precise predictions \cite{Aurenche:1985yk,Bern:2002,Balazs:2007hr} 
for this process in the theory in order to discover the Higgs boson as well 
as to unravel physics beyond the standard model.  Recently, unparticle physics 
\cite{Georgi:2007ek,Georgi:2007si} has gained a lot of attention 
in the context of collider signatures aiming to study new physics 
signals.  In a recent paper \cite{Kumar:2008zz}, it was found that 
the diphoton production at the hadron colliders is an important 
process to look for effects coming from unparticle physics.
This study was based on a leading order (LO) computation in QCD.  
Since the gluons contribute even at LO in the unparticle scenario, 
the higher order QCD effects are expected to play a significant role in 
the predictions.  In this paper we report contributions 
to next-to-leading order (NLO) in QCD, coming from a model 
with tensor unparticles as well as from the SM.
 
Parton level cross sections beyond the LO in perturbation theory encounter
IR singularities resulting from the soft gluons and massless collinear 
partons.  
In addition to the above singularities resulting from partons, the photons 
in the final states can also give collinear singularities when they become 
collinear to one of the final state partons.  These collinear singularities, 
often called QED singularities, can be avoided by appropriately constructing 
observables that are insensitive to them.  In addition, a parton in the 
final state can fragment non-perturbatively into a photon and a bunch of 
hadrons collinear to it, parameterized by photon fragmentation functions 
\cite{Bourhis:1997yu}.  The QED singularities can be removed by either 
absorbing them into these fragmentation functions or by suitable definition 
of diphoton events that removes these contributions.  The approach involving 
fragmentation    functions is theoretically tedious and also brings in 
additional non-perturbative input that is poorly known to date.  Hence we 
adopt an alternate smooth cone isolation criterion proposed by Frixione 
\cite{Frixione:1998jh} 
which ensures that the fragmentation contribution and the final state QED
singularities are suppressed.  

\section{Unparticle physics}

In a recent paper, Georgi \cite{Georgi:2007ek,Georgi:2007si} considered 
a scenario wherein the SM is weakly coupled to a sector which is scale 
invariant below an IR scale $\Lambda_{\cal U}$.  Consequently conventional 
particle interpretation at low energies in this new sector fails.  These 
fields were termed as unparticles by Georgi.  The unparticles could be 
interpreted as fractional number of massless particles \cite{Georgi:2007ek} 
or a collection of particles with a particular mass distribution 
\cite{Stephanov:2007ry} as a result of deconstruction.  
The coupling of these unparticles to the SM would lead to direct 
phenomenological consequences due to the peculiar unparticle phase 
space \cite{Georgi:2007ek} and propagator \cite{Georgi:2007si,
hg2-kingman}, which are determined by scale invariance.  Their 
coupling to the SM Higgs 
could lead to a constraint on the unparticle sector due to breaking of 
the scale invariance when the Higgs gets a vacuum expectation value
\cite{susy}.  As a consequence unparticle physics may be relevant
at the high energy colliders which have been actively investigated
\cite{Georgi:2007ek,Georgi:2007si,hg2-kingman,Mathews:2007hr,
Kumar:2008zz,collider}.  
In \cite{Grinstein:2008qk}, the corrected propagator and lower bounds on
the scaling dimensions of the vector and the tensor operators were obtained 
using
unitarity constraints on scattering amplitudes.  In this paper, we use
these new corrected results to do our phenomenology.

The unparticle operators could be of scalar, vector, tensor or fermionic 
type.  In the present article we restrict ourselves to tensor unparticles 
\cite{Mathews:2007hr,Kumar:2008zz} coupled to SM fields given by
\begin{eqnarray}
{\lambda_t \over \Lambda^{d_{\cal U}}_{\cal U}} T_{\mu \nu} ~ 
O^{\mu \nu}_{\cal U}\,,
\label{eq1}
\end{eqnarray}
where $\lambda_t$ is the dimensionless coupling constant for the unparticle 
tensor operator $O^{\mu \nu}_{\cal U}$ which is traceless and symmetric and 
has a scaling dimension $d_{\cal U}$.  $T_{\mu \nu}$ is the energy momentum 
tensor of the SM. Scale invariance restricts the scaling dimension of
tensor unparticle operator to $d_{\cal U} \geq 3$ 
\cite{Grinstein:2008qk}.
Conformal invariance on the other hand leads to a constraint $d_{\cal
U}\ge 4$ on the second rank tensor operators.  Scale and conformal
symmetries can only guide us on fixing the tensor structures of the
propagator leaving the overall normalisation undetermined. Unlike the
conformal invariance, the scale invariance does not fix the relative
coefficients of the tensors appearing in the tensor propagator
\cite{Grinstein:2008qk}.  We use the following propagator for our
phenomenological study:
\begin{eqnarray}
\int d^4x ~ e^{i k\cdot x}
\langle 0 |
     T O^{\mu \nu}_{\cal U}(x) O^{\alpha \beta}_{\cal U}(0)
|0 \rangle 
  &=&
    -i 
       C_T \frac {  \Gamma(2 -d_{\cal U})} {4^{d_{\cal U}-1} 
       \Gamma(d_{\cal U}+2)}
       ( - k^2)^{d_{\cal U}-2} 
\nonumber\\ 
  &&  \times  \left[ d_{\cal U}(d_{\cal U}-1)
        (g_{\mu \alpha} g_{\nu \beta} + \mu \leftrightarrow \nu)
      + \ldots \right] \, .
\label{prop}
\end{eqnarray}
where we have chosen $C_T$ =1.  The terms given by ellipses do not contribute 
to the diphoton production.
The terms in the ellipses depend on tensors, proportional to $g^{\mu \nu}$, 
$k^\mu$ and $k^\nu$.  The exact tensorial form of course depends on the 
symmetry (scale or conformal).  
These terms do not contribute to physical processes thanks to the 
conservation and traceless nature of the SM energy momentum tensor.  
Hence the symmetry restriction enters only through the 
scaling dimension $d_{\cal U}$ (the overall undetermined constant 
could be different for the scale and conformal invariant propagators).  
Hence we can safely use the above propagator Eq.~(\ref{prop}) with 
$d_{\cal U} \ge 3$ ($\ge 4$) for scale (conformal) invariant analysis.  
As larger $d_{\cal U}$ values give smaller unparticle contributions, 
we would demand only scale invariance which allows smaller values of 
$d_{\cal U}$.  In addition, as the SM energy momentum tensor is a 
conserved quantity, it does not require any operator mixing under SM 
renormalisation.

For our study, we closely followed Georgi's approach \cite{Georgi:2007ek}
to fix the coefficient of the effective interaction term at the weak scale 
for our phenomenology.  Accordingly, the effective interaction at the weak 
scale is proportional to
\begin{eqnarray}
C_{\cal U} \frac{\Lambda_{\cal U}^{d_{BZ} -d_{\cal U}}}{ M_{\cal U}^k} \, ,  
\end{eqnarray}
where $M_{\cal U}$ is the mass of heavy mediators in the hidden sector, 
$\Lambda_{\cal U}$ is the scale at which dimensional transmutation occurs 
and $C_{\cal U}$ is the Wilson coefficient.  $d_{BZ}$ and $d_{\cal U}$ 
are the scaling 
dimensions of Banks-Zaks ($BZ$)\cite{BaZa} and the unparticle operators 
respectively.  In the 
case of tensor unparticle operator coupled to the energy momentum tensor 
of the SM fields (having scaling dimension 4), the combination 
$C_{\cal U} \Lambda_{\cal U}^{d_{BZ}}/M_{\cal U}^k$ becomes dimensionless, 
which we call
$\lambda_t$ in Eq.\ \ref{eq1}.  This can be phenomenologically 
interesting only if we assume the dimensional transmutation scale 
closer to the electroweak scale and hence we choose $\Lambda_{\cal U}$ to 
be of the order of TeV.

\section{Diphotons at hadron colliders}

The hadronic cross section, $d\sigma^{P_1,P_2}$, in the QCD improved 
parton model can be expressed in terms of mass factorized parton level 
subprocess cross sections, $\Delta_{ab}$ convoluted with PDFs $f_{a/P}$ 
and $f_{b/P}$:
\begin{eqnarray}
d\sigma^{P_1,P_2}(S)=\sum_{a,b=q,\overline q,g}\int~dx_a\int~dx_b 
~ f_{a/P_1}(x_a,\mu_F^2) ~ f_{b/P_2}(x_b,\mu_F^2) ~ 
\Delta_{ab}(x_a,x_b,S,\mu_F^2)
\, ,
\label{QPM}
\end{eqnarray}
where $x_a$, $x_b$ are the momentum fractions of the incoming partons 
$a$ and $b$ respectively, $S$ is the invariant mass square of the
hadronic system and $\mu_F$ is the factorization scale.  Various 
kinematic constraints on the final states can be applied by introducing 
delta functions of these constraints.  
At LO in QCD  we have contributions of order
$\alpha^2, \alpha\lambda_t^2, \lambda_t^4$.
Here $ \alpha $ is the fine structure constant.
At this order, the diphoton signal in the SM 
comes from the process
$q +\overline q \rightarrow \gamma+\gamma$
and in the unparticle physics we have 
$q +\overline q \rightarrow \gamma+
\gamma$ and $g +g \rightarrow \gamma+\gamma$.
The presence of the gluon initiated process at LO in QCD is due to the fact 
that
the unparticle fields couple to quarks and gluons with the same strength.
At NLO, contributions are of the order 
$\alpha_s\alpha^2, \alpha_s \alpha \lambda_t^2$, and $\alpha_s \lambda_t^4$.
In the SM, 
$q +\overline q \rightarrow \gamma +\gamma + {\rm one~ loop}$,
$q +\overline q \rightarrow \gamma +\gamma + g$
and 
$q(\overline q) + g \rightarrow \gamma +\gamma +q (\overline q)$
contribute.
In the unparticle case, in addition to the above processes,
$g + g \rightarrow \gamma +\gamma + {\rm one~ loop}$ and
$g + g \rightarrow \gamma +\gamma + g$,
also contribute.
In addition, at the NLO SM $ gg \rightarrow \gamma \gamma $
box diagram contributes via its interference with leading
order $gg \rightarrow \gamma \gamma $ diagram with an unparticle 
propagator.

Before we compute the observable with the desired final state kinematical 
constraints, we compute the reduced cross sections ($\Delta_{ab}$ in
Eq.\ (\ref{QPM})) that are free of both IR and ultraviolet singularities.  
The UV divergences resulting from the processes involving loops are regulated 
using dimensional regularization and renormalized in the ${\overline {MS}}$ 
scheme.  

To study various kinematical distributions of these diphoton events with 
the experimental cuts, a fully 
analytical computation at the NLO level is tedious.  Hence we have opted 
for a semi-analytical approach, namely, a two cutoff phase-space slicing 
method 
\cite{Harris:2001sx}.  Here, the phase space integrals are split into
regions that are sensitive to soft and collinear singularities and 
those that are free of them.  Such a slicing of the phase space is 
done using two very small parameters denoted by $\delta_s$ and $\delta_c$ which 
define the boundaries of soft and collinear regions respectively.  
Processes involving real gluons or quark/antiquark emissions 
can be decomposed as  
\begin{eqnarray}
d\sigma^{real}_{ab}= 
d\sigma^{real}_{ab,soft}(\delta_s) +d\sigma^{real}_{ab,col}(\delta_s,\delta_c)
+d\sigma^{real}_{ab,fin}(\delta_s,\delta_c) \, .
\end{eqnarray}
The subscripts $col$ and $fin$ correspond to collinear and finite parts 
respectively.  In the above equation, $d\sigma^{real}_{ab,soft}(\delta_s)$ 
is the cross section wherein the phase space integrals of the out going 
gluons are constrained such that the energy of the gluon in the CM frame 
of the incoming partons, is in the range $0 \le E_g \le \delta_s 
\sqrt{s_{ab}}/2$.  $s_{ab}$ is the invariant mass squared of the incoming 
partons $a$ and $b$.  
Similarly, $d\sigma^{real}_{ab,col} (\delta_s,\delta_c)$ is the cross 
section where the collinear regions of the final state partons 
are integrated out keeping the invariant mass 
squared of the pair of collinear partons to be less than $\delta_c s_{ab}$.  
The remaining cross section where the soft and collinear regions are 
removed is denoted by $d\sigma^{real}_{ab,fin} (\delta_s,\delta_c)$ and 
can be evaluated in 4-dimensions itself.  The integrals involving soft 
and collinear regions of the phase space are regulated using dimensional 
regularization and can be computed analytically.  The next step involves 
the computation of the virtual gluon corrections denoted by $\sigma^V_{ab}$ 
to the born processes to NLO 
in QCD.  The soft and collinear singularities present in the loop integrals 
of these virtual corrections are regulated using dimensional regularization.  
The soft and collinear singularities 
coming from the phase space as well as from the loop integrals manifest 
themselves as 
poles in $\varepsilon$.  One can easily show that the following combination
\begin{eqnarray}
d\sigma^{S+V}_{ab}(\delta_s) \equiv d\sigma^{real}_{ab,soft}(\delta_s)
+\sigma^V_{ab} 
\, ,
\end{eqnarray}
is free of $\varepsilon$ poles coming from the soft gluons.   The collinear 
singularities from the $\sigma^V_{ab}$ and $d\sigma^{real}_{ab,col}
(\delta_s,\delta_c)$ will go away if the mass factorization counter terms 
denoted by $d \sigma^F_{ab}$, defined in the ${\overline {MS}}$ scheme, are 
included in the cross sections.  The combination
\begin{eqnarray}
d\sigma^{S+V+C+F}_{ab}(\delta_s,\delta_c)=d\sigma^{S+V}_{ab}(\delta_s)+
                    d\sigma^{real}_{ab,col}(\delta_s,\delta_c)+d \sigma^F_{ab}
\, ,
\end{eqnarray}
is free of all IR singularities in QCD.
Finally we end up with IR finite cross section given by
\begin{eqnarray}
\Delta_{ab}=d\sigma^{S+V+C+F}_{ab}(\delta_s,\delta_c)+d\sigma^{real}_{ab,fin}
(\delta_s,\delta_c)\, .
\label{deltaab}
\end{eqnarray}
Even though $d \sigma^{S+V+C+F}_{ab}$ and $d \sigma^{real}_{ab,fin}$ are 
separately dependent on $\delta_s$ and $\delta_c$, their sum is expected to 
be independent of these parameters.  Compact and simple analytical 
expressions for soft and collinear parts of the cross sections are 
reserved for a detailed publication \cite{future}.  

The reduced cross section Eq.\ (\ref{deltaab}) is IR finite.
The additional final state QED singularities coming from the 
SM subprocess $q (\bar q) ~g \to \gamma \gamma ~q (\bar q)$
are suppressed by the smooth cone isolation criteria advocated by Frixione 
\cite{Frixione:1998jh}.
The isolation criteria on each of the photons is imposed by using the longitudinal 
boost invariant dimensionless parameter 
$R_{i\gamma}=\sqrt{(\eta_i-\eta_\gamma)^2 +(\phi_i-\phi_\gamma)^2}$ where 
$\eta_i$ and $\eta_\gamma$ are pseudo-rapidities of the outgoing hadron 
$i$ and the photon respectively.  Similarly, $\phi_i$ and $\phi_\gamma$ are 
their azimuthal angles with respect to the beam direction.  Within a circle 
of radius $R_{0}$ centered at each photon 
we reject any event unless the following condition is 
fulfilled:
\begin{eqnarray}
\sum_i E_{T,i}~ \theta(R-R_{i\gamma}) \le {\cal H}(R)\, , \quad \quad 
{\rm for ~~all} ~~ R \le R_{0} \, ,
\end{eqnarray}
where $E_{T,i}$ is the transverse energy of the hadron $i$.  
The function ${\cal H}(R)$ must 
vanish as $R$ goes to zero in order to get IR finite observable.  
One such choice put forth by
Frixione is
\begin{eqnarray}
{\cal H}(R)= E^{iso}_{T} \left({1-\cos (R)\over 1-\cos(R_{0})}\right)^n
\, ,
\end{eqnarray} 
where $E^{iso}_{T}$ is a fixed energy and $n \ge 1$. 

Since the isolation criteria on diphoton events does 
not allow any hard partonic (hadronic) activity closer (collinear) to
either of the  photons, the fragmentation contributions which are collinear 
in nature are suppressed.  The isolation criteria also ensures the cancellation 
of the soft singularities coming from the real and virtual gluons as it does 
not restrict the phase space of the soft gluons.  By imposing finite transverse 
momentum cuts on the isolated photons, the smooth cone isolation criteria 
does not reject any hard parton which could be collinear to the initial 
states--- ensuring mass factorization.  Hence the isolation criteria on 
the diphotons guarantees the use of the phase space slicing method to compute 
an IR safe observable.   In addition, we can safely employ the experimental 
cuts 
on the rapidity $|y^\gamma|$ of the photons. 
\section{Results}
In this section we present our results for $\sqrt{S}=14$ TeV at the LHC.
The virtual and real corrections have been evaluated in the limit of vanishing
quark masses. 
We have evaluated LO and NLO cross sections with CTEQ6L and CTEQ6M parton
density sets \cite{Pumplin:2002vw} 
with the corresponding strong coupling constant $\alpha_s(M_Z)=0.118$.
The fine structure constant is $\alpha=1/128$.  
A single scale $Q$, the invariant mass of the
photon pair, is used for the renormalization and
factorization scale. 
The scale $\Lambda_{\cal U}$ at which scale invariance sets 
in the BZ sector is chosen to be 1 TeV. Scale invariance restricts 
the scaling dimension of tensor operators to $ d_{\cal U} \geq 3$,
in our calculation we have chosen $ d_{\cal U} =3.01$.  
The coupling $\lambda_t$ is taken to be of order one.

We have imposed the kinematical cuts on the 
photons as used by the ATLAS detector 
\cite{atlas}:  
 $p_T^\gamma >40~ (25) $ GeV for harder (softer) photon,
 $|\eta_\gamma| < 2.5$ for each photon. 
The photons are restricted to have a separation
of at least $R_{\gamma \gamma}=0.4$. In addition
to these constraints, we impose smooth cone 
isolation criterion on photons given in Eq.\ (8).
$R_0=0.4$ is taken for the cone radius.
Unless otherwise specified we use $ n=2$ and $ E^{iso}_T =15$ GeV
which appear in ${\cal H}(R)$
\begin{figure}[htb]
\centerline{
\epsfig{file=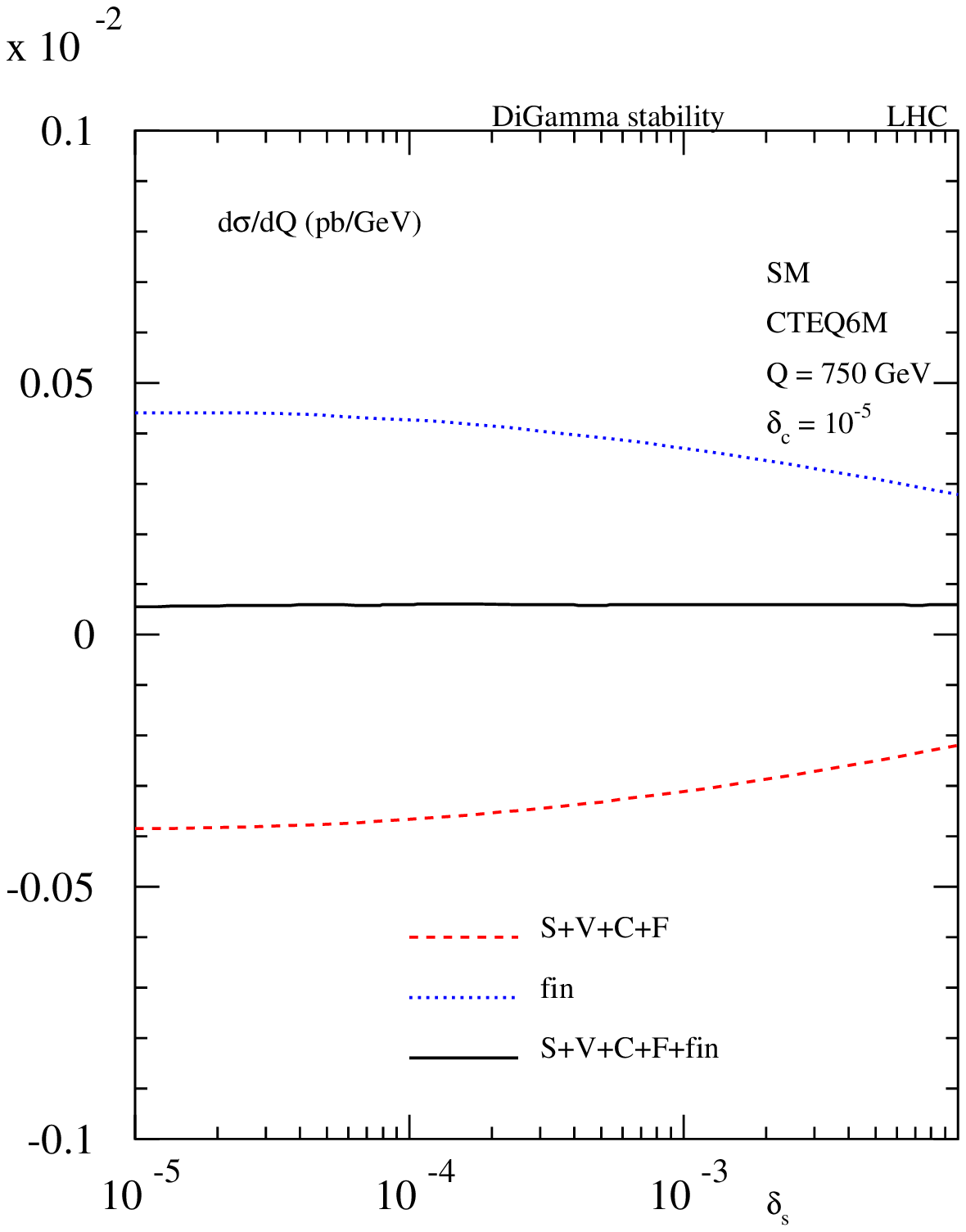,width=8cm,height=9cm,angle=0}
\epsfig{file=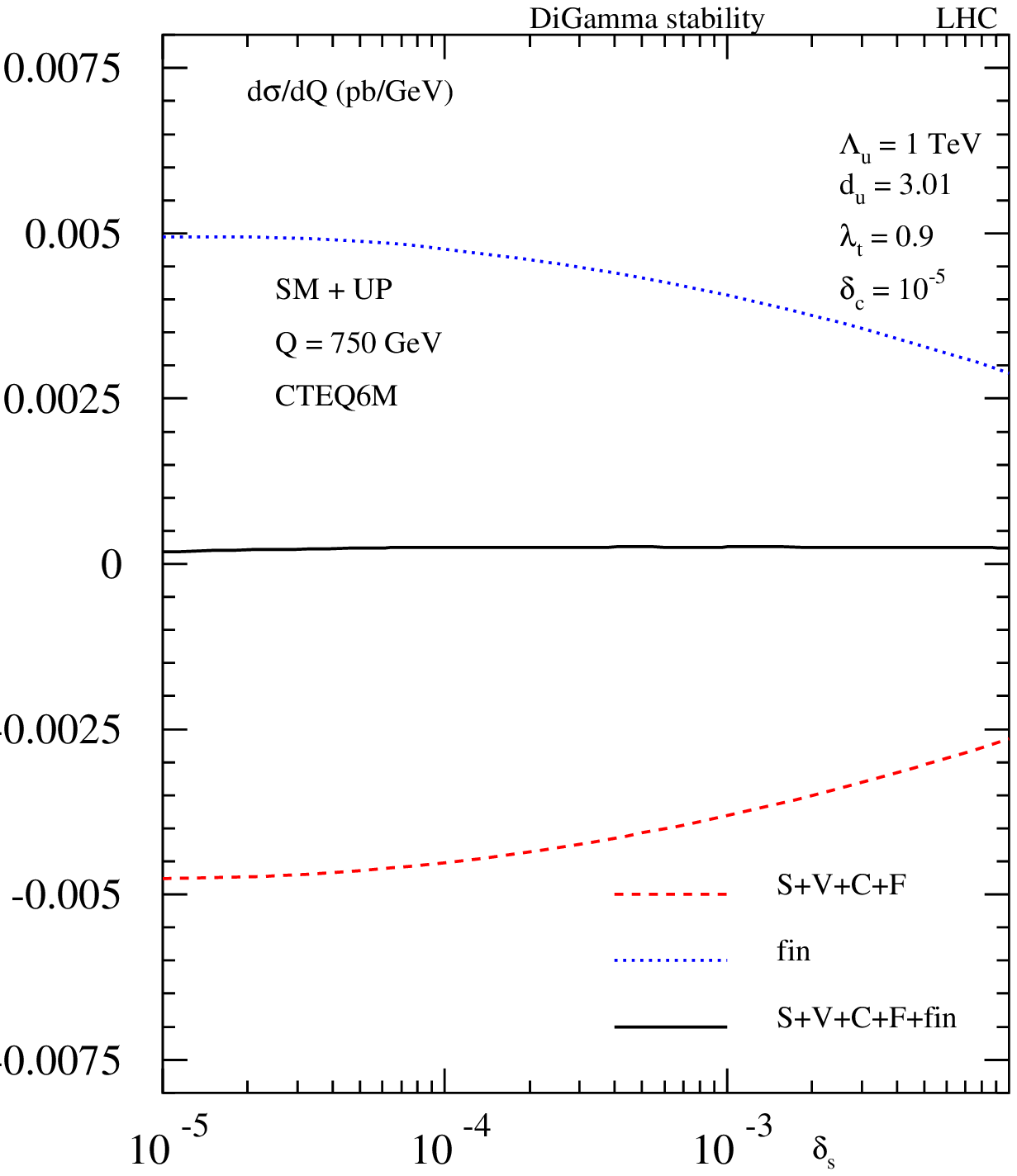,width=8cm,height=9cm,angle=0}}
\caption{Plots showing stability of $d\sigma/dQ$ for the SM (left panel) and 
the SM+ UP (right panel) against $\delta_s$ variation with the choice of 
$\delta_c=10^{-5}$.}
\label{ds}
\end{figure}

\begin{figure}[htb]
\centerline{
\epsfig{file=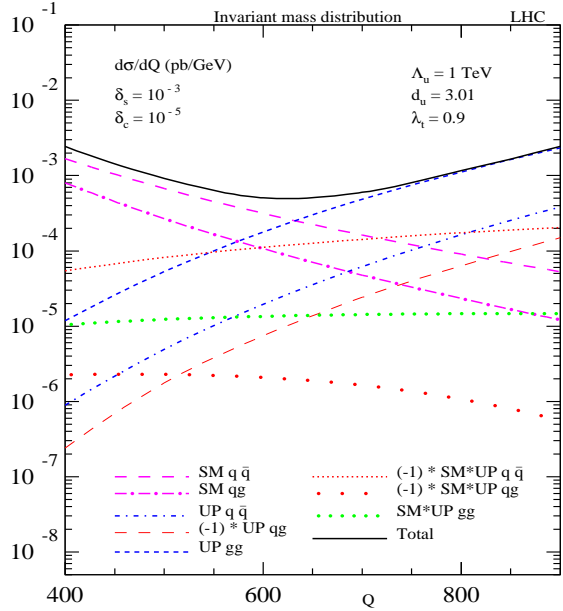,width=8cm,height=9cm,angle=0}
}
\caption{
Subprocess contributions in the SM and unparticle model in
the invariant mass distribution at NLO for $400 < Q <900$ GeV.}
\label{qfsub}
\end{figure}

\begin{figure}[htb]
\centerline{
\epsfig{file=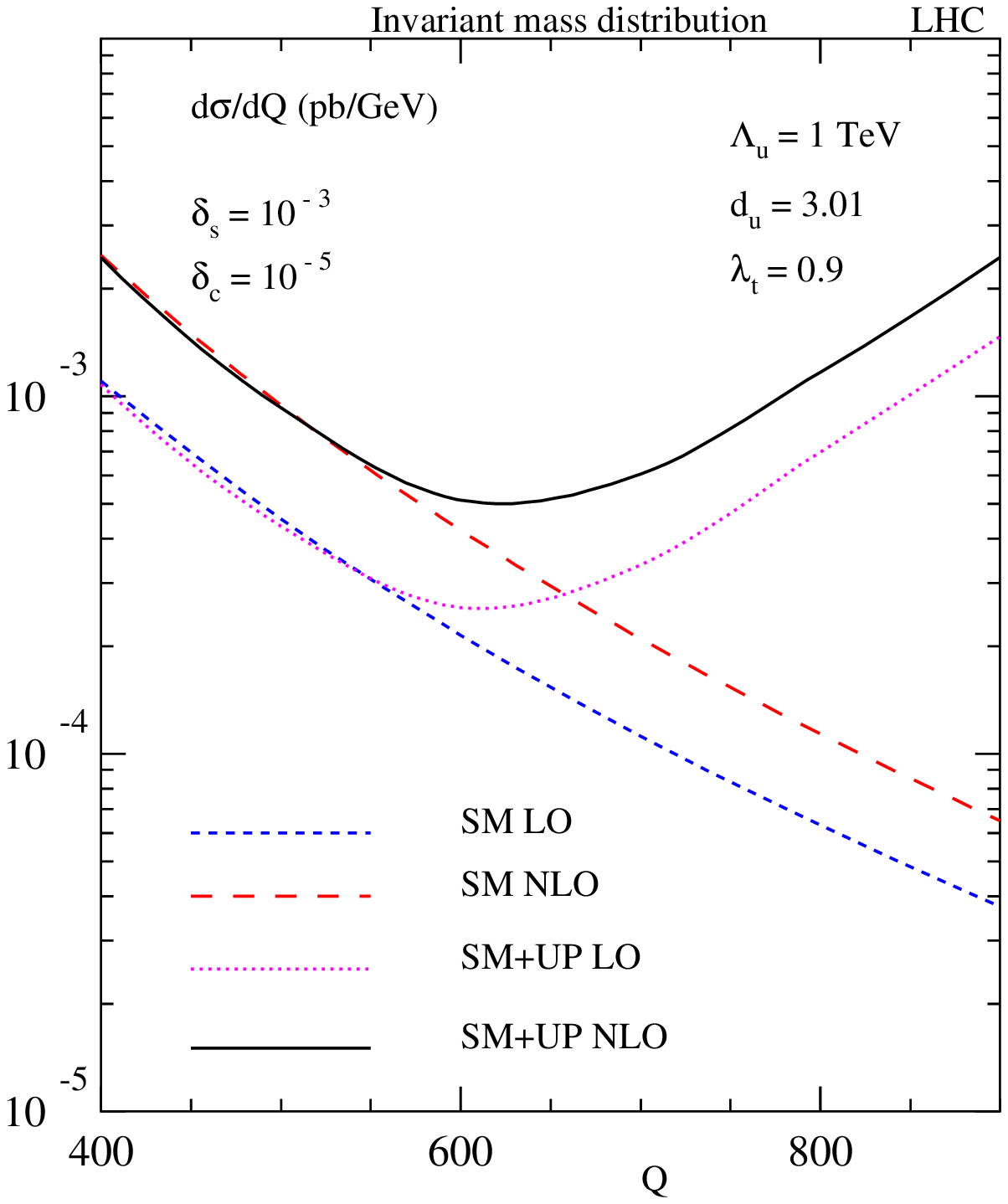,width=8cm,height=9cm,angle=0}
\epsfig{file=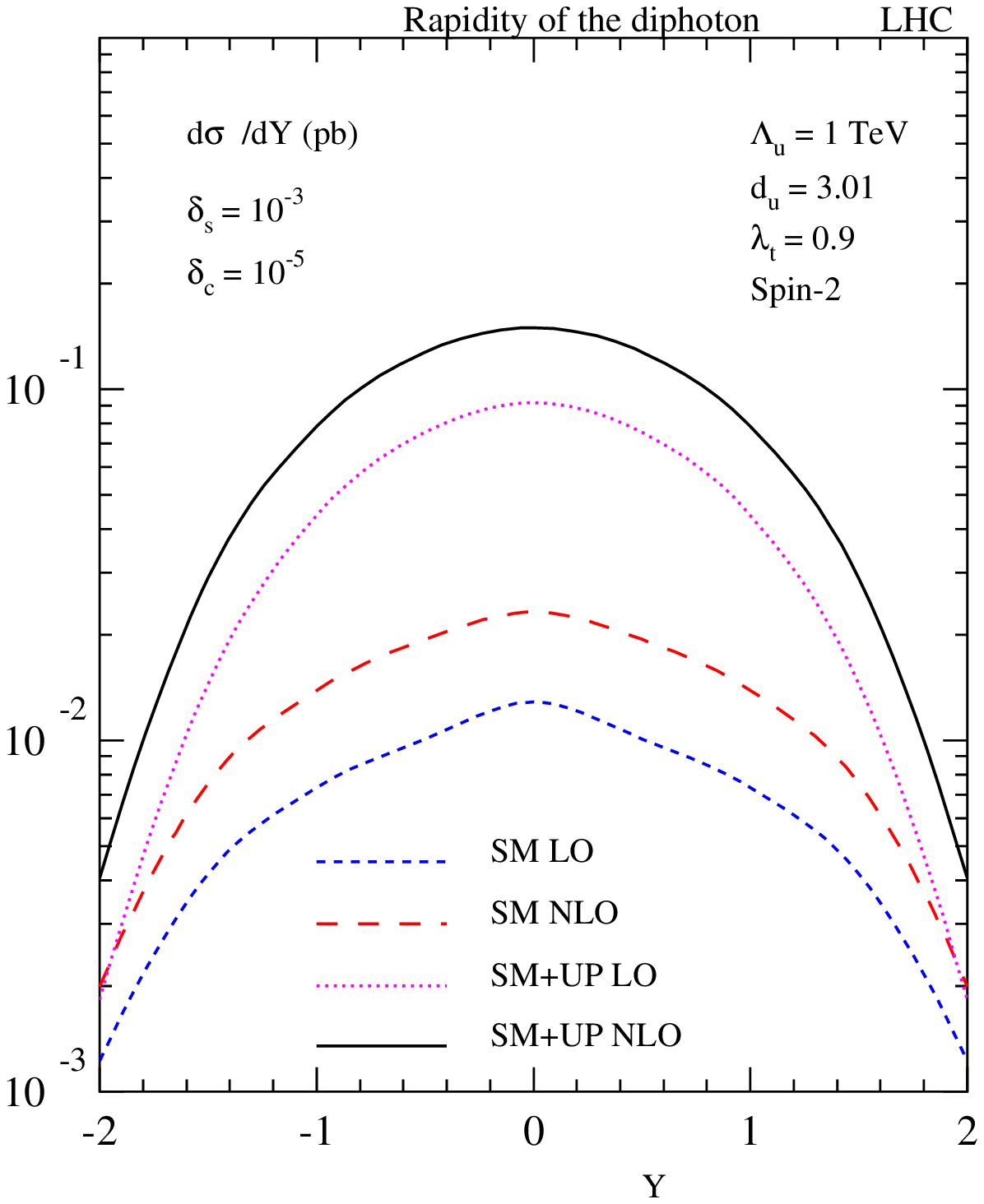,width=8cm,height=9cm,angle=0}
\label{sub}
}
\caption{Plots showing invariant mass (left panel) and rapidity 
(right panel) distributions of the diphoton system with 
$d_{\cal U}=3.01$, $\Lambda_{\cal U}=1$ TeV and $\lambda_{t}=0.9$.
For rapidity distribution $Q$ is integrated in the range $600$ GeV 
$< Q < 0.9 \Lambda_{\cal U}$.}
\end{figure}

\begin{figure}[htb]
\centerline{
\epsfig{file=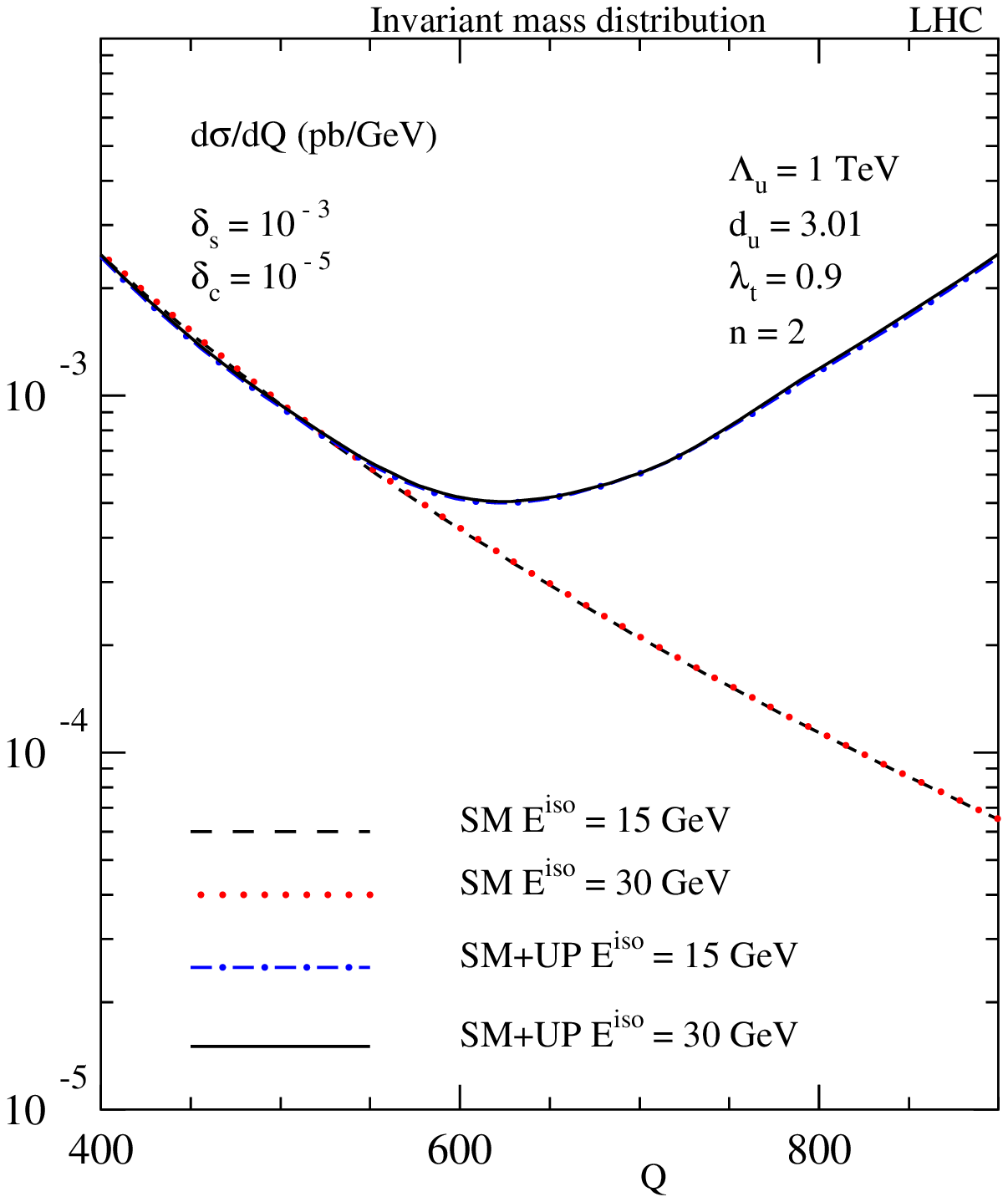,width=8cm,height=9cm,angle=0}
\epsfig{file=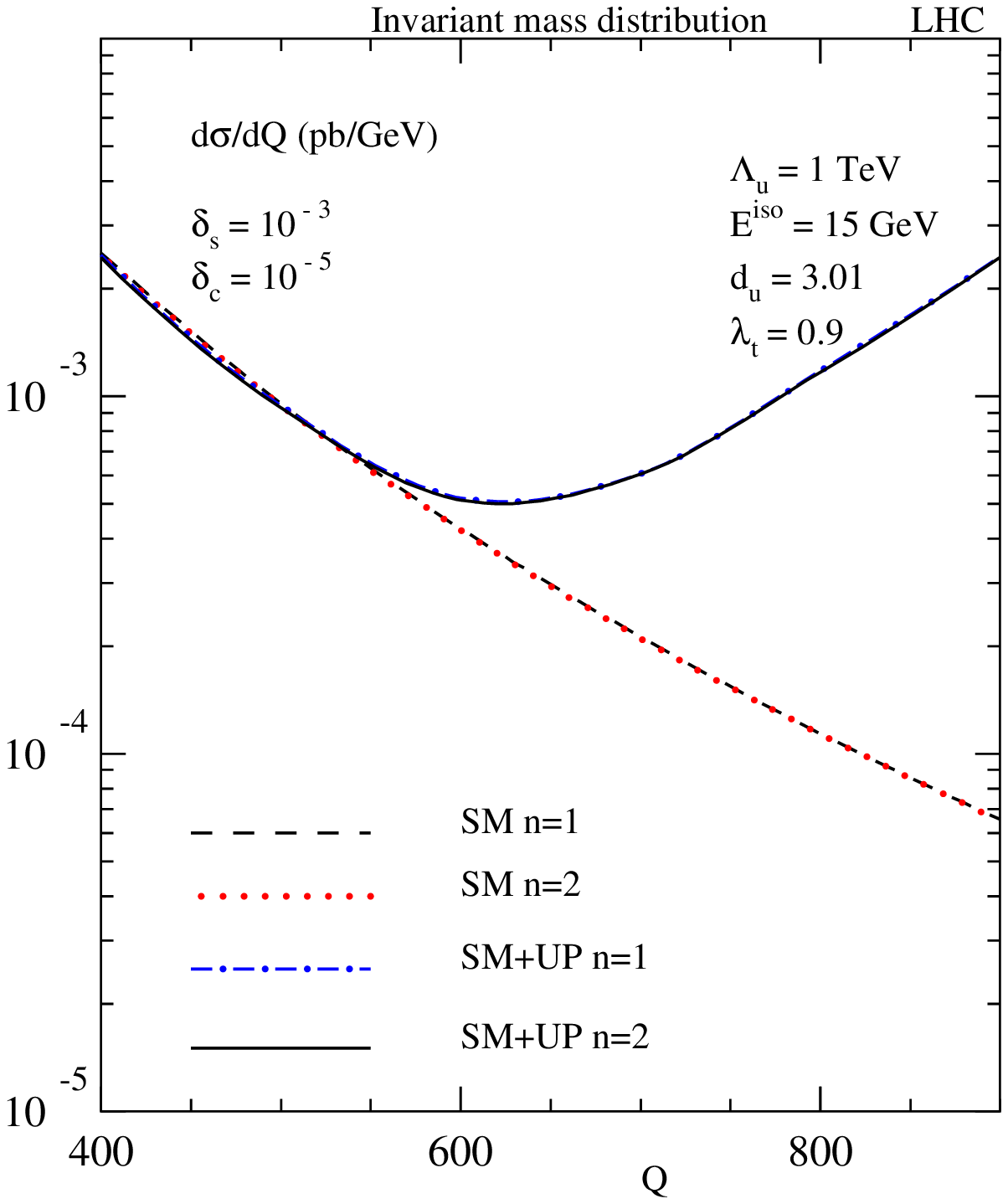,width=8cm,height=9cm,angle=0}
\label{sub1}
}
\caption{
Dependence on isolation parameters $ E^{iso}_T $ (left panel)
and $ n$ (right panel) is shown in invariant mass distribution
for the range $ 400 < Q < 900 $ GeV. 
}
\end{figure}

To show that the results are 
independent of the choice of 
the slicing parameters 
$\delta_s$ and $\delta_c$, 
we have plotted
$d\sigma_{ab}^{S+V+C+F}(\delta_s,\delta_c)$ 
and $d\sigma^{real}_{ab,fin}(\delta_s,\delta_c)$  
pieces of $d\sigma/dQ$ as functions of 
$\delta_s$ with $\delta_c $ kept fixed at a very small value $10^{-5}$
in Fig.\ \ref{ds}. 
We see that the sum is fairly stable under the variation.
The variation is around 2 \% for SM 
and is less then 5 \% for the signal (SM+UP).
For rest of our numerical study, we have chosen $\delta_s = {10}^{-3}$ and 
$\delta_c=10^{-5}$.
As SM NLO results exist in literature 
\cite{Aurenche:1985yk,Bern:2002,Balazs:2007hr}, to further check
our code we have compared our results with \cite{Bern:2002} using 
their isolation criterion [$ {\cal H}(R)= p_T(\gamma) \epsilon
               \left({[1-\cos (R)]/[1-\cos(R_{0})]}\right)^n $]
and their choice of $\mu_F$, $\mu_R$ and PDFs.  We have found
good agreement with \cite{Bern:2002}.

We present various subprocess contributions to NLO in QCD
in the invariant mass distribution for the range $ 400 < Q < 900 $ GeV
in Fig.\ \ref{qfsub}.
In the SM both $q\bar{q}$ and $qg$ subprocess contributions are positive with
$q\bar{q}$ contribution being dominant over that of $qg$ for the range
of $Q$ considered.  However for smaller values of Q($<150$ GeV), $qg$ 
contribution will be dominant due to the large gluon flux.
In the unparticle sector the  $q\bar{q}$ and $gg$ subprocess contributions 
via the pure unparticle exchange (direct) are positive while the $qg$ 
contribution is negative. 
At the interference level the $gg$ interference with the SM box has a
positive contribution and is almost constant for the range of $Q$ considered.  
However, the interference of both $q\bar{q}$ and $qg$ subprocesses 
with the SM have negative contributions and are larger in magnitude
compared to the direct ones.
As a result, the total ${\cal U}$-particle 
contributions (direct+interference) can have a pattern that will change 
the sign at some high value of $Q$.  Consequently, below this $Q$  
the signal can be lower than the SM background.
The direct  ${\cal U}$-particle exchange contributions can 
become significant in the large $Q$ region because the cross sections
go as powers of 
$Q/\Lambda_{\cal U}$, thus leading to the visibility of 
the ${\cal U}$-particles only in that region of $Q$. 
It is worth noting that only the $gg$ initiated subprocess
has the dominant contribution over the rest in this region.

\begin{figure}[htb]
\centerline{
\epsfig{file=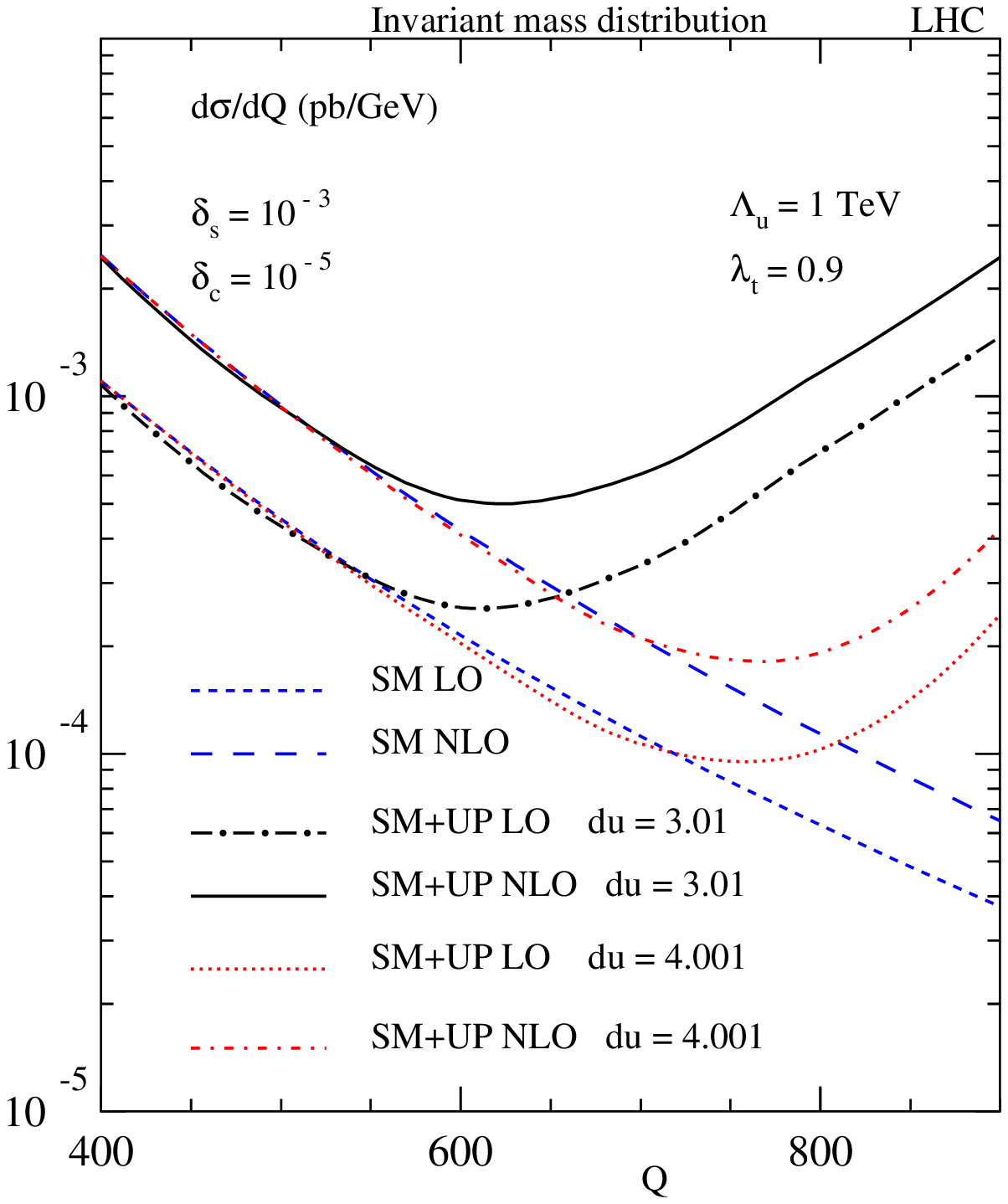,width=8cm,height=9cm,angle=0}
\epsfig{file=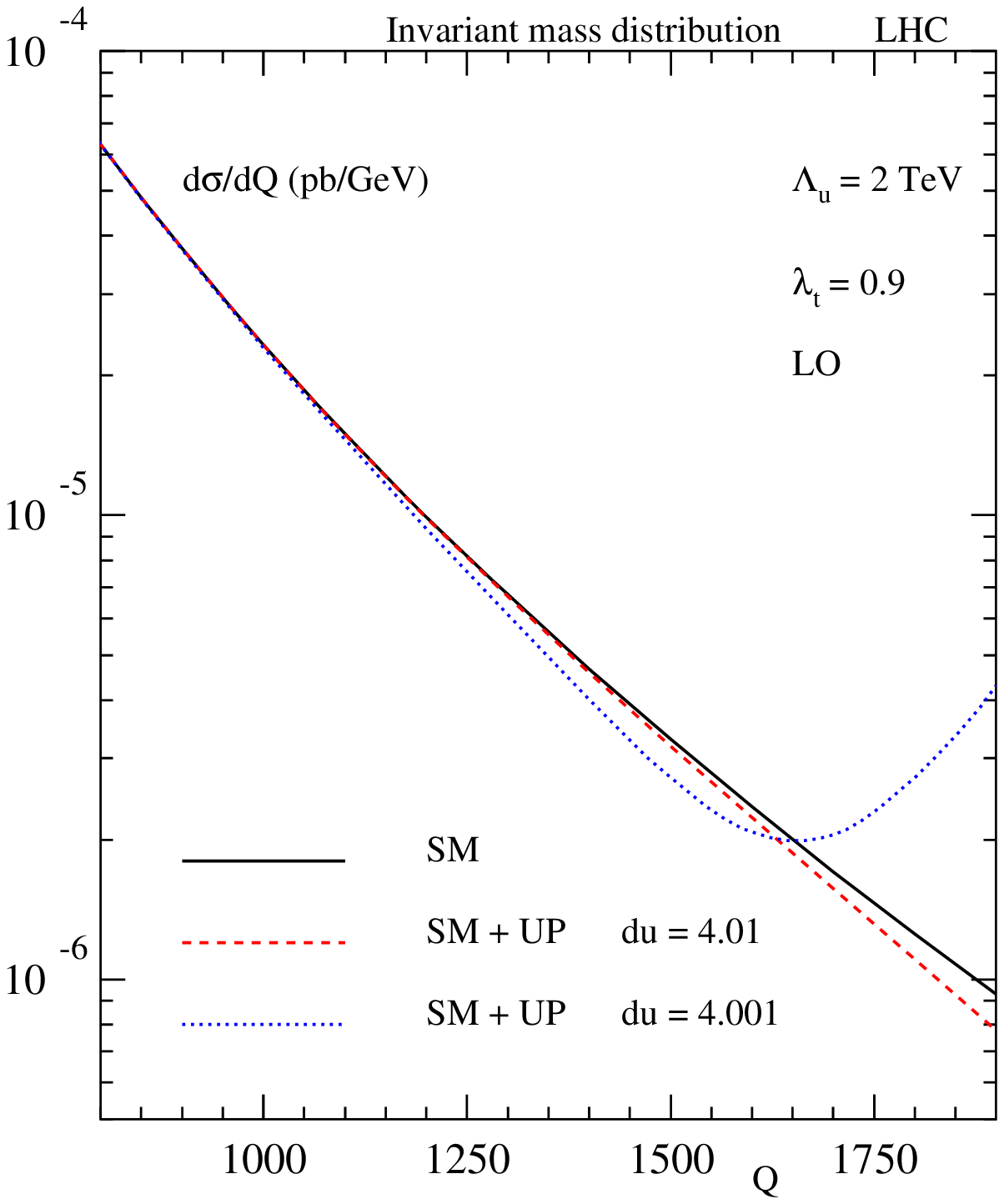,width=8cm,height=9cm,angle=0}
}
\caption{Invariant mass distribution of the diphoton system (left panel) 
for $d_{\cal U}>4$ (conformal invariance) is contrasted with the 
$3< d_{\cal U} < 4$ (scale invariance).  To LO in QCD and for 
$\Lambda_{\cal U}=2$ TeV (right panel), we have plotted the invariant 
mass distribution of the diphoton system for $d_{\cal U}>4$.
}
\label{dugt4}
\end{figure}

In Fig.\ 3, we present our results for $d \sigma/dQ$ (left panel) for 
$Q$ between $400$ and $900$ GeV and $d\sigma/dY$ (right panel) for 
$|Y|< 2.0$, where $Y$ is the rapidity of the diphoton system.  
The short dashed (LO) and the long dashed lines (NLO) correspond to SM 
distributions. The dotted (LO) and the solid (NLO) lines correspond to 
the signal (SM+UP) of the unparticle physics.
As expected we find that the unparticle 
contribution to the invariant mass distribution grows
with $Q$ and it dominates over the SM contributions above $Q=600$ GeV.
The precise value where this happens will depend very much on the choice
of $\Lambda_{\cal U}$ and other unparticle parameters.
Since unparticle effects can be seen in the larger values of $Q$, 
rapidity distributions are computed  by integrating $Q$ between 
600 and 900 GeV. 
Near the central value of $Y$, we find large enhancement of the 
cross section from the SM results if we include unparticle contributions.

At next-to-leading order the total transverse energy of the hadrons 
is due to a single parton around the photons and does
not correspond to the actual hadronic energy in an experiment.
Hence the $E^{iso}_T$ at the parton level is a crude 
estimate of that of the jets at the detector level.
To show the dependence on 
$E_T^{iso}$ we present the  invariant mass distribution
for $E_T^{iso} = 15 $ GeV and $E_T^{iso} = 30 $ GeV,
for $n=2$.
To study the dependence of our predictions 
on the choice of ${\cal H}(R)$ we have varied it by changing $n$ 
from $1$ to $2$ and keeping $E_T^{iso}$ fixed at a value of $15$ GeV.
These variations are shown in Fig.\ \ref{sub1}
and show a very small dependence for $R_0 = 0.4$.

Until now our analysis was restricted to the case where $3< d_{\cal U} < 4$,
which was essential for a tensor unparticle as a consequence of scale 
invariance.  There are no known examples of unitary quantum field theory
that are scale invariant but not conformal invariant.  For conformal
invariance, unitarity demands that $d_{\cal U} > 4$ for the tensor unparticles.
In Fig.~\ref{dugt4} (left panel) the unparticle sector as a result of 
scale invariance and not conformal ($d_{\cal U} =3.01$) is contrasted to the 
case where the unparticle sector is conformal ($d_{\cal U} =4.001$).  This is
to both LO and NLO in QCD and for $\Lambda_{\cal U}=1$ TeV.
In Fig.~\ref{dugt4} (right panel) we have considered the invariant mass 
distribution
for $d_{\cal U} > 4$ to LO in QCD.  For this plot we have considered 
$\Lambda_{\cal U} = 2$ TeV and have probed $Q < 0.9 \Lambda_{\cal U}$.  Closer to 
$d_{\cal U}=4$ there could still be sufficient unparticle contribution for 
the tensor operator.  The turn-around with energy behaviour of the 
unparticle effects is a typical feature of any physics beyond the SM.  
One observes a similar behaviour in models with large extra-dimensions.  
The origin of this turn around contribution comes from the terms 
proportional to $(s/\Lambda_{\cal U})^{d_{\cal U}}$ in the matrix elements involving 
unparticles.
Finally the observable we have considered 
for the diphoton process does not distinguish other beyond 
standard model contact interactions.  

\section{Conclusions}
In this article we have used one of the important channels namely the diphoton 
production in order to explore the scale invariant unparticle sector coupled 
to the SM.   We have restricted ourselves to contributions coming from the tensor 
unparticles.  Since the QCD plays an important role at the LHC, we have 
systematically included all the partonic subprocesses that enter at the 
NLO level in the strong coupling constant $\alpha_s$.  We have employed smooth 
cone isolation criteria advocated by Frixione in order to reduce
fragmentation contributions and thereby removing final state QED singularities.
We have applied appropriate cuts on the photon kinematics that are used by the 
ATLAS Collaboration.  Since it is hard to perform a fully analytic computation 
with isolation criteria and experimental cuts on the final state photons, 
we have resorted to the two cut-off 
phase space slicing method.  We have also shown that our results are 
insensitive to the slicing parameters that bear no physics but are introduced 
in the intermediate stages of the computation.  Our SM results
are in agreement with the ones in the literature.
We have presented two important
distributions, namely, the invariant mass and the rapidity distributions of the 
diphoton system with the above isolation criteria and the relevant experimental 
cuts. 
For $\Lambda_{\cal U} = 1$ TeV, we have presented 
various subprocess contributions to the signal in 
the invariant mass distribution and found that the
$ gg $ contribution is dominant at high $Q$ values
where the unparticle effects are visible.
Both in the SM and in the unparticle case, the NLO contributions resulting from $\alpha_s$ enhance
the cross sections, thanks to the large gluon flux at the LHC.  Our NLO results 
to the diphoton production including the unparticle effects not only estimate the 
corrections from QCD to the leading contributions but also reduce the scale 
uncertainties coming from the factorization scale \cite{future}.
We have also considered the case where the unparticle sector
is a result of scale invariance and not conformal ($3< d_{\cal U} < 4$) and also 
the case where it is due to conformal invariance ($d_{\cal U} >4$) for the tensor
unparticle operator.

\vspace{.5cm}
\noindent
{\bf Acknowledgments:}  We thank B.W.~Harris for providing his code in 
which the phase-space slicing method is implemented.  MCK would like to 
thank CSIR, New Delhi for financial support.  The work of VR and AT has 
been partially supported by funds made available to the Regional Centre 
for Accelerator-based Particle Physics (RECAPP) by the Department of 
Atomic Energy, Govt.  of India.  AT and VR would like to thank the  
cluster computing facility at Harish-Chandra Research Institute where 
a part of the computational work for this study was carried out.  We thank 
the referee for all the suggestions.

\end{document}